\documentclass{jfm}
\usepackage[utf8]{inputenc}
\usepackage{graphicx}
\usepackage{subcaption}
\usepackage{newtxtext}
\usepackage{newtxmath}
\usepackage{enumitem}
\usepackage[normalem]{ulem}
\usepackage{xcolor}
\usepackage{natbib}
\usepackage[export]{adjustbox}
\usepackage{hyperref}
\hypersetup{
    colorlinks = true,
    urlcolor   = blue,
    citecolor  = black,
}

\newcommand{\RomanNumeralCaps}[1]
\def\la{\left<}
\def\ra{\right>}

\def\bnabla{\boldsymbol{\nabla}}

\def\lb0{\ell_*|_{\beta_*=0}}

\def\LS{{\text{LS}}}
\def\SS{{\text{SS}}}

\title{Quantitatively mapping the Eady model onto a two-layer quasi-geostrophic model}

\author{J. Meunier, B. Gallet}
\affiliation{Université Paris-Saclay, CNRS, CEA, Service de Physique de l’Etat Condensé, 91191 Gif-sur-Yvette, France.}

\begin{document}
\maketitle

\begin{abstract}

The two-layer quasigeostrophic model (2LQG) and the Eady model are two idealized systems illustrating the baroclinic instability of atmospheric jets and ocean currents. 
The two setups share many ingredients -- background vertically sheared zonal flow of density-stratified fluid in a rapidly rotating frame -- while differing in complexity and dimensionality. The Eady model has a continuous vertical direction, with baroclinic turbulence induced by {\it boundary} potential vorticity (PV) gradients at top and bottom. By contrast, the 2LQG sytem typically models baroclinic instability induced by {\it interior} PV gradients. This distinction challenges our ability to clearly identify a couple of `modes' through which the Eady dynamics could be inferred from a simpler 2LQG system.

In the present study, we show that this difficulty can be circumvented in the turbulent regime arising for weak bottom drag. Namely, guided by the common organization of both systems into a gas of coherent vortices, we identify a quantitative mapping between the Eady and the 2LQG models. The mapping allows for parameter-free predictions of the eddy diffusivity of the Eady model based on the knowledge of the 2LQG diffusivity. We illustrate these results using  numerical simulations of the Eady and 2LQG models with linear or quadratic bottom drag. 

\end{abstract}


\section{Introduction}

In rapidly rotating density-stratified fluids, large-scale vertically sheared zonal flows are in thermal-wind balance with a lateral buoyancy gradient. This configuration, common to both ocean currents and atmospheric jets, is prone to baroclinic instability and results in a turbulent flow. To model this phenomenon and assess the effective transport properties of such `baroclinic turbulence', one typically resorts to idealized models that retain only the key features of the system while avoiding unnecessary complexity. The two simplest such models are arguably the two-layer quasi-geostrophic (2LQG) model and the Eady model. The 2LQG model involves two fluid layers of different density lying on top of one another, and the dynamics is described using two streamfunctions characterizing the vertically invariant horizontal flow in each layer \citep{phillips1954, salmon1978two,salmon1980baroclinic, larichev1995, arbicfierl2004a, thompsonyoung2006,kohl2022diabatic,pudig2025baroclinic,sterl2025joint}. The governing equations of the 2LQG model are thus effectively two-dimensional. In contrast with the 2LQG model, the Boussinesq Eady model possesses a continuous vertical direction. All fields are functions of the three spatial coordinates, and time \citep{eady1949}. In the rapidly rotating limit, however, quasi-geostrophy (QG) allows for a crucial simplification. The potential vorticity (PV) vanishes in the interior of the domain, the consequence being that the dynamics reduce to 2D evolution equations for the streamfunctions at the top and at the bottom of the domain. 

Even in the simplest $f$-plane configuration considered in the present study, there are some crucial distinctions between the Eady and the 2LQG models. The Eady model is driven by boundary PV gradients at top and bottom, while the 2LQG model possesses interior PV gradients. This leads to different phenomenologies for the two models at horizontal scales much smaller than the Rossby deformation radius: the Eady model exhibits the small-scale fronts and filamentary structures characteristic of surface quasi-geostrophy \citep{constantin1994formation,scott2014numerical,Lapeyre2017,valade2025surface}, whereas such structures are absent from the 2LQG model. From an analytical perspective, the distinction between boundary versus interior PV gradients leads to difficulties to isolate a unique set of vertical `modes' through which the Eady model could be projected onto an equivalent 2LQG model. Among the various attempts are the early method proposed by \cite{flierl1978}, the pedagogical approach of  \cite{vallis2017atmospheric}, who crudely approximates the Eady model using finite-differences with only two levels in the vertical, thereby obtaining a 2LQG model, and more general methods leading to non-unique bases of acceptable modes \citep{lapeyre2009vertical,smith2013surface,scott2012assessment,lacasce2012surface,tulloch2009quasigeostrophic,yassin2022discrete}. Retaining only two modes from such a basis typically leads to a 2LQG model that disagrees quantitatively with the original Eady model (e.g. model A in \cite{rocha2016galerkin}), especially if the models are to be compared over a broad range of parameter values.

Despite these difficulties, however, the close mathematical structures of the Eady and 2LQG models suggest that a quantitative mapping between the two systems may be within reach. In the present study we unveil such a mapping procedure by focusing on the low-drag regime. Besides showing that the large-scale dynamics of the two models are intimately connected, the mapping provides a way of quantitatively inferring the eddy diffusivity of the turbulent Eady model (among other quantities) based on the known behavior of the simpler 2LQG model. We conclude in section~\ref{sec:conclusion}.

We introduce the Eady model and the 2LQG model in section~\ref{sec:models}. The first step of the mapping procedure consists of a reduction of the 2LQG equations to retain only the terms that contribute to the low-drag asymptotic behavior. This task is carried out in section~\ref{section:reduced_eq}. The second step of the mapping procedure consists of a large-scale expansion of the Eady model, carried out in section~\ref{sec:Eadyexp}. The expansion shows that up to a rescaling, both models are governed by the same asymptotic set of equations in the low-drag regime. As illustrated in section~\ref{sec:mapping} for quadratic drag and in section~\ref{sec:linear} for linear drag, this allows for the quantitative and parameter-free prediction of the eddy diffusivity of the Eady model, based on existing scaling theories for the 2LQG diffusivity.

\section{The quasigeostrophic Eady and two-layer models\label{sec:models}}

\subsection{The Eady model}

\begin{figure}
    \centering
\includegraphics[width=\linewidth]{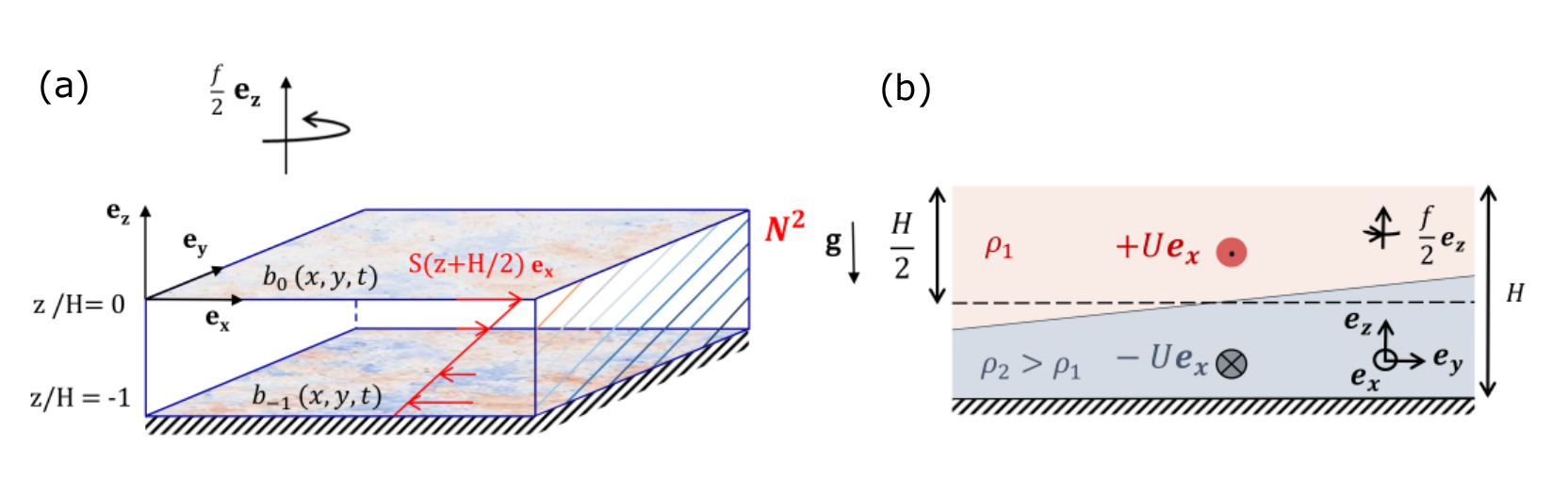}
    \caption{Two idealized models for the study of horizontally homogeneous baroclinic turbulence on the $f$-plane. \textbf{(a)} The Eady model features a continuous vertical direction $z$. 
    \textbf{(b)} By contrast, the two-layer quasi-geostrophic model consists of two stacked shallow fluid layers. In both models equilibrated turbulence results from a balance between potential energy release from the thermal-wind-balanced base state and dissipation by bottom drag.} 
    \label{fig:schema_eady_2LQG}
\end{figure}

The Eady model is sketched in Fig. \ref{fig:schema_eady_2LQG}(a). Consider a layer of fluid in a 3D domain $(x, y, z) \in [0, L] \times [0, L] \times [-H, 0]$, with weak density variations around a reference density $\rho_0$, described within the Boussinesq approximation. The fluid layer is subject to gravity ${\bf g}$ and rapid rotation at a rate $f/2$ around the vertical axis ${\bf e}_z$. The base state of the system consists of a vertically sheared zonal flow ${\bf U}(z)=S(z+H/2) {\bf e}_x$, with a uniform shear $S$. The density of the fluid is stably stratified along the vertical, with a uniform buoyancy frequency $N$. Because of thermal-wind balance, the vertically sheared flow ${\bf U}(z)$ implies the existence of a meridional buoyancy gradient $-fS$. That is, the background buoyancy field reads $B(y,z)=N^2 z - f S y$. 

We consider departures from this base state obeying periodic boundary conditions in the horizontal directions. For small enough isopycnal slope $fS/N^2$, the evolution of the system is conveniently described within the QG approximation. The extensive derivation of the QG approximation for the present setup is detailed in \cite{Gallet_2022}, and we only recall the main steps here. At lowest order in the QG expansion, the dominant flow satisfies geostrophic balance: the lowest-order flow is horizontal and divergence-free, with a streamfunction proportional to the departure pressure field. Hydrostatic balance indicates that the buoyancy departure is directly proportional to the vertical derivative of the departure pressure field. In particular, the dimensionless evolution equations for the buoyancy field at top and bottom read:
\begin{eqnarray}
\partial_t b_0-\partial_x p_0 + \frac{1}{2} \partial_x b_0 + J(p_0,b_0) & = & 0 \, , \label{eq:Eadyb0}\\
\partial_t b_{-1}-\partial_x p_{-1} - \frac{1}{2} \partial_x b_{-1} + J(p_{-1},b_{-1}) & = & \mu_\text{Eady} \bnabla \cdot \left( |\bnabla p_{-1} | \bnabla p_{-1} \right) \, , \label{eq:Eadybminus1}
\end{eqnarray}
where $p_0(x,y,t)$ and $b_0(x,y,t)$ denote the dimensionless pressure and buoyancy departures at the top of the domain, while $p_{-1}(x,y,t)$ and $b_{-1}(x,y,t)$ denote the dimensionless pressure and buoyancy departures at the bottom of the domain. In equations~(\ref{eq:Eadyb0}-\ref{eq:Eadybminus1}) and in the remainder of this subsection, we have non-dimensionalized time $t$ with $Sf/N$, horizontal coordinates $x$ and $y$ with the Rossby deformation radius $NH/f$, vertical coordinate $z$ with $H$, departure pressure with $\rho_0 H^2 S N$ and buoyancy with $H S N$.

Strictly speaking, equation~(\ref{eq:Eadybminus1}) corresponds to the evolution equation for buoyancy {\it near} the bottom of the domain, just above the small Ekman boundary layer. The rhs term thus represents advection of the background stratification by the vertical velocity arising from Ekman pumping inside the boundary layer. We consider a stress on the rough Ocean floor that is quadratic in outer flow velocity. The resulting pumping velocity is proportional to the curl of this quadratic stress, hence the rhs term in~(\ref{eq:Eadybminus1}), where $\mu_\text{Eady}$ denotes the dimensionless quadratic drag coefficient.

In the interior of the domain, the QG approximation naturally leads to material conservation of the quasi-geostrophic potential vorticity (QGPV). Because this equation has no source terms in the present context, one can focus on the vanishing QGPV state. Demanding that the QGPV be zero inside the domains readily provides a relation between the pressure at top and bottom and its vertical derivative (i.e. the buoyancy) at top and bottom, see \cite{Gallet_2022}. The resulting inversion relation is more conveniently expressed in spectral space, relating the boundary values of buoyancy and pressure (or streamfunction) for each horizontal wavevector ${\bf k}$:
\begin{eqnarray}
\left( \begin{matrix}
\hat{b}_0|_{\bf k}  \\
\hat{b}_{-1}|_{\bf k} 
\end{matrix} \right) & = &
\left[ \begin{matrix}
\frac{k}{\tanh (  k)} &  -\frac{ k}{\sinh (k)}  \\
\frac{ k}{\sinh (k)} & -\frac{ k}{\tanh (k)}
\end{matrix} \right]
\left( \begin{matrix}
\hat{p}_0|_{\bf k}  \\
\hat{p}_{-1}|_{\bf k} 
\end{matrix} \right) \, , \label{eq:inversionEady}
\end{eqnarray}
where $\hat{\phi}|_{\bf k}$ denotes the Fourier coefficient of the field $\phi$ and the horizontal wavevector ${\bf k}$ is non-dimensionalized using the Rossby deformation radius $NH/f$.

The QG Eady system consists of equations~(\ref{eq:Eadyb0}-\ref{eq:Eadybminus1})  with the inversion relation~(\ref{eq:inversionEady}). We are interested in the long-time statistics of solutions to this system in a doubly periodic domain $(x,y)\in [0,L]^2$, with large-enough domain-size $L$ for these statistics to be independent of $L$. A central quantity of interest is the meridional buoyancy flux induced by the turbulent flow in statistically steady state. One easily shows that this buoyancy flux is depth-independent for the present vanishing-PV states \citep{Gallet_2022,GRLMeunier}. The time- and volume-averaged buoyancy flux is thus readily obtained from the time- and horizontal area-average of the meridional buoyancy flux evaluated at the top or at the bottom of the domain. Denoting as $D_{\text{Eady}}$ the buoyancy flux non-dimensionalized with $NH^2S/f$, we simply obtain:
\begin{eqnarray}
D_{\text{Eady}}=\la \overline{b_0 \partial_x p_0}  \ra=\la \overline{b_{-1} \partial_x p_{-1}}  \ra  \, , \label{eq:defDeady}
\end{eqnarray}
where the angular brackets denote space average and the overbar denotes time average.

\subsection{The 2LQG model}

The two-layer quasi-geostrophic (2LQG) model corresponds to further simplification as compared to the Eady model. As for the Eady model, we only discuss the main characteristics of the 2LQG model, redirecting the interested reader to standard textbooks for explicit derivations \citep{salmonbook,pedlosky2013,vallis2017atmospheric}. The model is sketched in Fig. \ref{fig:schema_eady_2LQG}(b). A shallow fluid layer of uniform density $\rho_1$ sits above a shallow fluid layer of density $\rho_2 > \rho_1$, both layers having equal depth at rest. The fluid layers are subject to gravity and uniform global rotation around the vertical axis at a rate $f/2$, where $f$ denotes the  Coriolis parameter. We focus on the limit of rapid rotation and shallow fluid layers, so that the system is governed by quasigeostrophic (QG) dynamics. The Rossby deformation radius of the 2LQG model is $\lambda= \frac{\sqrt{g'H}}{2f_0}$, with $g' = \frac{\rho_2 - \rho_1}{\rho_1} g$ the reduced gravity.

We denote as ${\bf u}_1(x,y,t)$ and ${\bf u}_2(x,y,t)$ the lowest-order velocity fields in the upper and lower layers, respectively.  The velocity fields ${\bf u}_1$ and ${\bf u}_2$ are horizontal, vertically-invariant and divergence-free in the horizontal plane. They derive from streamfunctions  $\psi_{1}(x,y,t)$ and $\psi_{2}(x,y,t)$, that is ${\bf u}_{1,2}=-\bnabla \times (\psi_{1,2} \, {\bf e}_z)$.
The base state consists of a vertically sheared zonal flow with uniform velocities $+U \mathbf{e}_x$ and $-U \mathbf{e}_x$ in the upper and lower layers, respectively. 
As a consequence of geostrophic balance and hydrostatic balance, the vertical displacement of the interface from its rest position is directly proportional to $\psi_{1}-\psi_{2}$. 
The vertically sheared base flow corresponds to $\psi_{1}-\psi_{2}=-2Uy$, and therefore the interface between the two layers is tilted in the meridional direction (See Fig. \ref{fig:schema_eady_2LQG}). This tilt of the interface corresponds to a uniform meridional gradient of vertically averaged buoyancy, in thermal-wind balance with the sheared zonal flow.

We consider arbitrary departures from the base state, expressed in terms of two fields $\psi(x,y,t)$ and $\tau(x,y,t)$ defined by $(\psi_{1}+\psi_{2})/2=\psi$ and $(\psi_{1}-\psi_{2})/2=-U y + \tau$. The field $\psi(x,y,t)$ is referred to as the `barotropic' streamfunction. It corresponds to the streamfunction of the vertically averaged flow (average over both layers). The field $\tau(x,y,t)$ is the departure `baroclinic' streamfunction. It encodes the departure of the vertically averaged buoyancy from its base-state uniform gradient. Because of this correspondence between $\tau$ and the vertically averaged buoyancy, $\tau$ is sometimes referred to as the buoyancy variable, even though it has the dimension of a streamfunction. Correspondingly, $U$ will sometimes be referred to as the background meridional buoyancy gradient in the following.

The QG expansion turns the shallow-water equations into conservation equations for the potential vorticities in both layers. The sum and the difference of these equations yield the coupled evolution equations for $\psi$ and $\tau$, known as the barotropic PV and baroclinic PV equations. Non-dimensionalizing space with $\lambda$ and time with $\lambda/U$, these equations read:
\begin{eqnarray}
\partial_t  \Delta \psi + J(\psi,\Delta \psi) + J(\tau,\Delta \tau) + \partial_x \Delta \tau   &  = &    {\cal D} \, ,  \label{eq:psi2LQG} \\
\partial_t  (\Delta \tau-\tau) + J(\psi, \Delta \tau - \tau) + J(\tau,\Delta \psi)   + \partial_x (\Delta \psi + \psi)    & = & -{\cal D}  \,  . \label{eq:tau2LQG}
\end{eqnarray}
We have included in the model a quadratic drag force in the lower layer, proportional to the squared lower-layer departure velocity. This drag term leads to the contributions $\pm {\cal D}$ on the rhs of (\ref{eq:psi2LQG}) and (\ref{eq:tau2LQG}), where:
\begin{eqnarray}
{\cal D} & = & -\frac{\mu_*}{2} \bnabla \cdot \left[ |\bnabla (\psi-\tau) | \bnabla(\psi-\tau) \right]\, ,  \label{eq:defdragD} 
\end{eqnarray}
with $\mu_*$ the quadratic drag coefficient non-dimensionalized with $\lambda$.

We are interested in the long-time statistics of solutions to (\ref{eq:psi2LQG}) and (\ref{eq:tau2LQG}) in a doubly periodic domain ${x,y}\in [0,L]^2$, with large-enough domain-size $L$ for these statistics to be independent of $L$. The main quantity of interest is the meridional buoyancy flux induced by the turbulent flow in statistically steady state. Non-dimensionalizing this flux with $U$ and $\lambda$ leads to the effective diffusivity $D_*=\la \overline{ \psi_x \tau} \ra$. When simulating this system numerically, for numerical stability we also include hyperviscous terms $-\nu \Delta^4(\Delta \psi )$ and $-\nu \Delta^4 (\Delta \tau - \tau)$ on the rhs of equations (\ref{eq:psi2LQG}) and (\ref{eq:tau2LQG}), respectively. We focus on the regime where the hyperviscosity $\nu$ is small enough for $D_*$ to be independent of $\nu$. Investigating the efficiency of the turbulent transport then boils down to  characterizing the dependence of the dimensionless diffusivity $D_*$ with the dimensionless drag coefficient $\mu_*$.

\subsection{Turbulent cascades in the 2LQG model}

\begin{figure}
    \centering
    \includegraphics[width=0.8\linewidth]{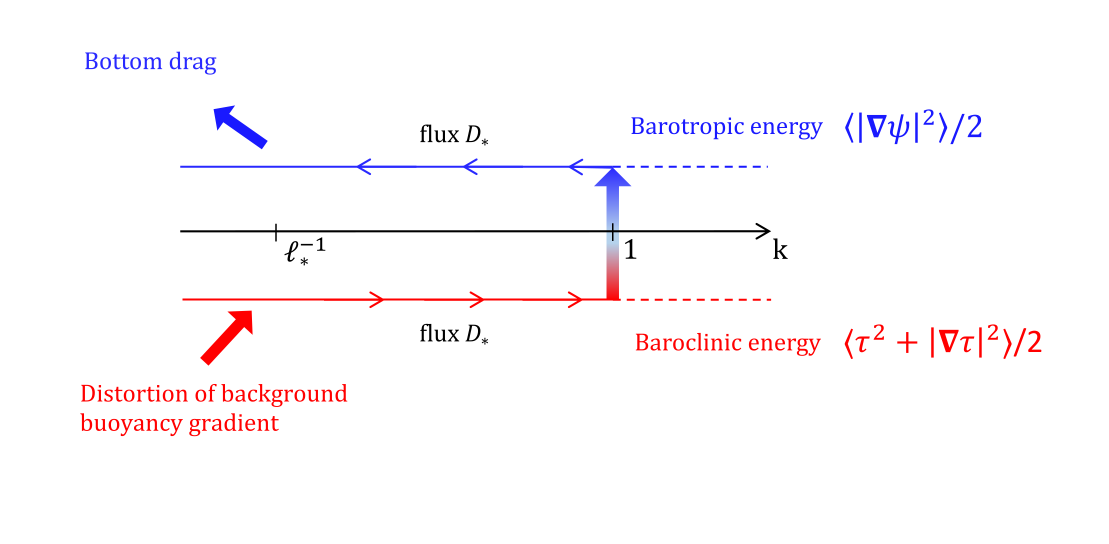}
    \caption{Energy cascades in baroclinic QG turbulence, as described in \cite{salmon1978two}. According to this scenario, energy conversion from the baroclinic mode to the barotropic mode arises near the Rossby deformation radius $\lambda$.}
    \label{fig:dual_cascade}
\end{figure}

In a similar fashion to the 2D Navier-Stokes (2DNS) equation, the 2LQG model features both an energy and an enstrophy invariant. The energy equation is obtained by multiplying equation~(\ref{eq:psi2LQG}) with $\psi$ and equation~(\ref{eq:tau2LQG}) with $\tau$, before summing the resulting equations and averaging over the domain:
\begin{eqnarray}
\frac{\mathrm{d}}{\mathrm{d}t} \left[ \frac{\la |\bnabla \psi|^2 \ra}{2} + \frac{\la |\bnabla \tau|^2 \ra}{2} +\frac{\la \tau^2 \ra}{2} \right] & = & \la \psi_x \tau \ra + \la (\tau - \psi) {\cal D} \ra \, , \label{eq:energy2LQG}
 \end{eqnarray}
where we omit the negligible contribution from hyperviscosity. The three terms inside the square brackets respectively correspond to barotropic kinetic energy, baroclinic kinetic energy and potential energy. Time-average leads to the energy power integral:
\begin{eqnarray}
D_* & = &  \la \overline{(\psi - \tau) {\cal D}} \ra \, , \label{eq:powerintegral}
 \end{eqnarray}
relating the rate of release of potential energy, $D_*= \la \overline{\psi_x \tau} \ra $, to the power dissipated by the drag term. The precise energy pathways in 2LQG turbulence are represented schematically in Fig. \ref{fig:dual_cascade}, adapted from the work of Salmon \citep{salmon1978two,salmon1980baroclinic}. When the drag coefficient is small, the flow develops large structures up to a typical scale $\ell \gg \lambda$, denoted as $\ell_*=\ell/\lambda \gg 1$ in dimensionless form. The large-scale flow distorts the background gradient of $\tau$ through the term $\partial_x \psi$ in equation~(\ref{eq:tau2LQG}). This process is a source of baroclinic potential energy, associated with the injection term $D_*=\la \overline{\psi_x \tau} \ra$  on the lhs of the energy power integral~(\ref{eq:powerintegral}). The input potential energy $\la \tau^2 \ra/2$ follows a forward cascade from the large scale $\ell$ (dimensionless scale $\ell_*$) down to smaller scales. However, as for standard 2DNS turbulence the conservation of both energy and enstrophy forbids the forward energy cascade from reaching all the way to the small hyperviscous scale. Instead, the forward cascade of baroclinic energy stops at scales comparable to the deformation radius (dimensionless scale $1$), where baroclinic energy is efficiently converted into barotropic kinetic energy. The latter then obeys an inverse cascade akin to the standard inverse energy cascade of the 2DNS equation. Barotropic energy is transferred all the way up to the large scale $\ell_*$ where drag efficiently dissipates it. The frictional sink in the power integral~(\ref{eq:powerintegral}) is thus dominated by the contribution from the barotropic flow,
\begin{align}
\nonumber \la \overline{(\psi - \tau) {\cal D}} \ra & = -\frac{\mu_*}{2} \la \overline{(\psi - \tau) \bnabla \cdot \left[ |\bnabla (\psi-\tau) | \bnabla(\psi-\tau) \right] }\ra \simeq -\frac{\mu_*}{2} \la \overline{\psi \bnabla \cdot \left[ |\bnabla \psi | \bnabla \psi \right] }\ra \\
& = \frac{\mu_*}{2}  \la \overline{ |\bnabla \psi |^3}\ra \, . \label{eq:simplediss}
\end{align}
The power integral thus reflects a balance between the rate of release of potential energy due the distortion of the background buoyancy gradient by the large-scale barotropic flow, and the rate of damping of barotropic kinetic energy through bottom friction.

\subsection{Vortex gas scaling theory}

The turbulent cascade phenomenology of Salmon provides invaluable insight into the qualitative behavior of the 2LQG system. However, it proves insufficient to make quantitative predictions for the turbulent transport properties of the system~\citep{chang_control_2019,chang_parameter_2021,chen_revisiting_2023}, namely for the scaling relation between $D_*$ and $\mu_*$. The reason is that the cascade theory is framed in spectral space, and therefore it misses any large-scale organization of the flow in physical space. As observed by~\citet{thompsonyoung2006}, however, the large-scale barotropic flow organizes into a dilute vortex gas, that is, an ensemble of vortices with a core radius much smaller than the typical distance $\ell_* \gg 1$ between two vortices. The coherent vortices are crucial because, even though the vortex cores occupy a small fraction of the area of the domain, they are responsible for most of the frictional dissipation~(\ref{eq:simplediss}) entering the power integral~(\ref{eq:powerintegral}).  
Such large-scale flow organization into a vortex gas is by no means a property of the 2LQG model only. Indeed, we recently showed that the same phenomenology arises in the standard 2DNS equation, with important consequences for the transport properties of the flow~\citep{meunier2025}.

\section{Reduced equations for 2LQG baroclinic turbulence at low drag \label{section:reduced_eq}}

Before mapping the Eady model onto the 2LQG model, a preliminary step is to reduce the latter model by retaining only the dominant terms controlling the low-drag dynamics. To identify the dominant terms of equations~(\ref{eq:psi2LQG}-\ref{eq:tau2LQG}) at the `large scale' $\ell_*$ and at the `small-scale' $1$ (corresponding to the deformation radius in dimensional form), we first provide estimates for the various fields and their derivatives. The estimates at scale $\ell_*$ are denoted with a subscript LS for `large-scale estimate', while the estimates at scale $1$ are denoted with a subscript SS for `small-scale estimate'.

\subsection{Estimating large-scale quantities in the 2LQG model: vortex-gas model}

In the present 2LQG context, the vortex-gas scaling theory was initially put forward in~\citet{gallet2020}, before being refined in~\citet{hadjerci2023} to capture the regime of very-low quadratic drag. 
For the sake of brevity, we only recall here the main scaling relations without justification, referring the interested reader to~\citet{gallet2020}, \citet{hadjerci2023} and~\citet{meunier2025} for the detailed scaling arguments. The barotropic flow consists of a gas of vortices, with typical inter-vortex distance $\ell_* \gg 1$ and core radius scaling as $\sqrt{\ell_*}$. The barotropic streamfunction has a typical magnitude $\psi_\LS \sim \ell_*^2$. Correspondingly, the typical velocity in the inter-vortex region is $(\bnabla \psi)_\LS \sim \ell_*$. The effective diffusivity $D_*$ of the flow scales as the product of this large-scale velocity with the large scale $\ell_*$ of the flow, which leads to $D_* \sim \ell_*^2$. The baroclinic streamfunction scales as $\tau_\LS \sim \ell_*$, as deduced from a slantwise free-fall argument in GF (for instance). Such fluctuations of $\tau$ are produced at the large scale $\ell_*$ through the distortion of the background gradient by the vortex-gas flow. These estimates are gathered in the left-hand column of Table~\ref{table:scalings}. We also include the scaling $\nabla_\LS \sim \ell_*^{-1}$ for the spatial derivatives at large scale, and a scaling of the time derivative based on the typical turnover time of the large-scale barotropic flow, $(\partial_t)_\LS \sim (\bnabla \psi )_\LS/\ell_* \sim 1$.

\subsection{Estimating small-scale quantities in the 2LQG model: Salmon's dual cascade theory}

We now consider spatial scales comparable to the deformation radius, referred to as `small scales' in the following, the deformation radius being much smaller than the integral scale of the flow (that is, $\ell_* \gg 1$ in dimensionless form). With our non-dimensionalization, such scales correspond to spatial derivatives of order $\nabla_\SS \sim 1$. Estimates for $\psi$ and $\tau$ at the deformation scale can be readily obtained from Salmon's dual cascade theory. For the inverse energy cascade of barotropic kinetic energy, a standard `Kolmogorov' estimate for the velocity increment at scale $r$ is $\delta (|\bnabla \psi|) \sim (D_* r)^{1/3}$, where $\delta$ denotes the two-point increment over a dimensionless separation $r$ and $D_*$ denotes the dimensionless energy flux (usually denoted as $\epsilon$ in the standard literature). Upon substituting $r=\ell_*$ and $r=1$ and taking the ratio of the resulting expressions we obtain
\begin{equation}
\frac{(\bnabla \psi)_\text{LS}}{(\bnabla \psi)_\SS} \sim \ell_*^{1/3} \Rightarrow (\bnabla \psi)_\SS \sim \ell_*^{2/3} \, ,  \label{eq:gradpsiSS}
\end{equation}
where we used  $(\bnabla \psi)_\LS \sim \ell_*$. Combining~(\ref{eq:gradpsiSS}) with $\nabla_\SS \sim 1$ yields $\psi_\SS \sim  \ell_*^{2/3}$.
In a similar fashion, the forward cascade of temperature variance $\tau^2$ leads to the following estimate for the temperature increment $|\delta \tau|  \sim (D_* r)^{1/3}$, where $D_*$ denotes the dimensionless flux of temperature variance (see figure~\ref{fig:dual_cascade}). Substituting $r=\ell_*$ and $r=1$ and dividing the resulting expressions leads to
\begin{equation}
\frac{\tau_\LS}{\tau_\SS} \sim \ell_*^{1/3} \Rightarrow \tau_\SS \sim \ell_*^{2/3} \, , \label{eq:gradpsiLS}
\end{equation}
where we have inserted $\tau_\LS \sim \ell_*$ to obtain the second expression. 
Finally, we estimate the time derivative of small-scale quantities based on their advection by the fast large-scale barotropic flow: $(\partial_t)_\SS \sim (\bnabla \psi)_\LS \cdot \bnabla_\SS \sim \ell_*$. The estimates for $\psi$, $\tau$ and their derivatives at small scale are reported in the right-hand column of Table~\ref{table:scalings}.

\begin{table}
    \centering
    \begin{tabular}{ccc}
      										& LS estimates  & SS estimates  \\
 \hline 
   $\bnabla$ 								& $\ell_*^{-1}$ 				& $1$ \\
   $\partial_t$ 								& $1$ 					& $\ell_*$ \\
    \hline 
    $\psi$ 									& $\ell_*^2$ 					& $\ell_*^{2/3}$ \\
   $\bnabla \psi$ 							& $\ell_*$ 					& $\ell_*^{2/3}$ \\
   $\Delta \psi$ 								& $1$ 				& $\ell_*^{2/3}$ \\
    \hline 
    $\tau$ 									& $\ell_*$ 					& $\ell_*^{2/3}$ \\
   $\bnabla \tau$ 							& $1$ 					& $\ell_*^{2/3}$ \\
   $\Delta \tau$ 								& $\ell_*^{-1}$ 				& $\ell_*^{2/3}$ \\
    \end{tabular}
    \caption{Estimates of the various fields and their space and time derivatives in the inter-vortex region, both at the large dimensionless scale $\ell_*$ (LS, left-hand column) and at the small dimensionless scale $1$ associated with the deformation radius (SS, right-hand column). The scalings are reported in terms of powers of the large scale $\ell_* \gg 1$. Scaling-laws in terms of the dimensionless quadratic drag coefficient are obtained by substituting the vortex-gas scaling prediction $\ell_* \sim \mu_*^{-2/3}$.}
    \label{table:scalings}
\end{table}

\subsection{Reduced equations}

Starting from the undamped 2LQG equations, the first step of the reduction consists in identifying the dominant terms at the beginning and at the end of the turbulent cascades (scales $\ell_*$ and $1$). By forming reduced equations that retain such dominant terms at both ends of the cascades, we should capture the right dynamics throughout the inertial ranges of the cascades. 

Let us proceed by considering first the undamped version of the $\tau$-equation:
\begin{equation}
\partial_t  (\Delta \tau-\tau) + J(\psi, \Delta \tau - \tau) + J(\tau,\Delta \psi)   + \partial_x (\Delta \psi + \psi)      =  0  \,  . \label{eq:tau2LQGundamped}
\end{equation}
At the large scale $\ell_*$, following table~\ref{table:scalings}  the various terms scale as follows:
\begin{eqnarray}
\partial_t  (\Delta \tau) \sim \ell_*^{-1} \, , & \qquad & \partial_t   \tau  \sim \ell_* \, , \\
J(\psi, \Delta \tau) \sim  \ell_*^{-1} \, , & \qquad & J(\psi, \tau) \sim  \ell_* \, , \\
J(\tau,\Delta \psi)  \sim  \ell_*^{-1} \, , & \qquad & \partial_x \Delta \psi \sim \ell_*^{-1} \, , \qquad  \partial_x  \psi  \sim \ell_* \, , 
\end{eqnarray}
meaning that only the terms $-\partial_t   \tau $, $-J(\psi, \tau)$ and $\partial_x \psi$  need be retained to describe the large-scale dynamics of $\tau$. This aligns well with the standard phenomenology~\citep{salmon1978two,salmon1980baroclinic,larichev1995}. Namely, the temperature field behaves as a passive scalar at large scales, with fluctuations induced by the distortion of the background temperature gradient by the barotropic flow~\citep{thompsonyoung2006,gallet2020}.

At the small dimensionless scale $1$ (deformation radius), following table~\ref{table:scalings}  the various terms are estimated as follows:
\begin{eqnarray}
\partial_t  (\Delta \tau - \tau) \sim \ell_*^{5/3} \, , & \qquad & J(\psi, \Delta \tau-\tau) \sim (\bnabla \psi)_\LS \bnabla(\Delta \tau-\tau)_\SS \sim \ell_*^{5/3}  \, , \\
J(\tau,\Delta \psi)  \sim  \ell_*^{4/3} \, , & \qquad & \partial_x (\Delta \psi + \psi) \sim \ell_*^{2/3}  \, , 
\end{eqnarray}
meaning that only the terms $\partial_t  (\Delta \tau - \tau)$ and $J(\psi, \Delta \tau-\tau)$ need be retained to describe the small-scale dynamics of $\tau$. 
Including in a single equation the dominant terms at large scale and the dominant terms at small scale leads to the following reduced undamped equation:
\begin{equation}
\partial_t  (\Delta \tau-\tau) + J(\psi, \Delta \tau - \tau) + \partial_x  \psi  =  0  \,  . \label{eq:tau2LQGreducedundamped}
\end{equation}

We now turn to the undamped version of the $\psi$-equation:
\begin{equation}
\partial_t  \Delta \psi + J(\psi,\Delta \psi) + J(\tau,\Delta \tau) + \partial_x \Delta \tau    =   0 \, .
\end{equation}
At the large scale $\ell_*$, following table~\ref{table:scalings}  the various terms scale as follows:
\begin{eqnarray}
\partial_t  \Delta \psi \sim 1 \, , & \qquad & J(\psi,\Delta \psi) \sim 1  \, , \\
J(\tau,\Delta \tau)   \sim  \ell_*^{-2} \, , & \qquad & \partial_x \Delta \tau  \sim \ell_*^{-2}  \, , 
\end{eqnarray}
meaning that only the terms $\partial_t  \Delta \psi$ and $J(\psi,\Delta \psi)$ need be retained to describe the large-scale dynamics of $\psi$. 
At the small dimensionless scale $1$ (deformation radius), following table~\ref{table:scalings}  the various terms are estimated as follows:
\begin{eqnarray}
\partial_t  \Delta \psi \sim \ell_*^{5/3} \, , & \qquad & J(\psi,\Delta \psi) \sim (\bnabla \psi)_\LS \bnabla \Delta \psi _\SS \sim   \ell_*^{5/3}  \, , \\
J(\tau,\Delta \tau) \sim (\bnabla \tau)_\SS \bnabla \Delta \tau _\SS  \sim \ell_*^{4/3} \, , & \qquad & \partial_x \Delta \tau  \sim \ell_*^{2/3}  \, ,  \label{eq:estimatesSStau}
\end{eqnarray}
The dominant terms lead to the 2D Euler equation for $\psi$:
\begin{eqnarray}
\partial_t  \Delta \psi + J(\psi,\Delta \psi) & = & 0 \, . \label{eq:Euler4psi}
\end{eqnarray}
By keeping the lowest-order terms of each equation, we have reduced the system to an unforced, undamped Euler flow~(\ref{eq:Euler4psi}) that distorts the background gradient of some passive scalar field $\Delta \tau - \tau$, see equation~(\ref{eq:tau2LQGreducedundamped}). Such an Euler equation for $\psi$, uncoupled to $\tau$, cannot capture the dynamics of the full 2LQG system. Instead, according to Salmon's phenomenology, the barotropic flow is forced at small-scale through conversion of energy from the baroclinic component $\tau$, and it is damped at large-scale by friction. We thus complement equation (\ref{eq:Euler4psi}) with the first next-order term at small-scale, $J(\tau,\Delta \tau)$ (see the estimates (\ref{eq:estimatesSStau})), together with the lowest-order expression of the large-scale friction term, ${\cal D}  \simeq  -\frac{\mu_*}{2} \bnabla \cdot \left( |\bnabla \psi | \bnabla \psi \right)$. The result of this procedure is the following reduced set of equations:
\begin{eqnarray}
\partial_t  \Delta \psi + J(\psi,\Delta \psi) + J(\tau,\Delta \tau)  &  = &    -\frac{\mu_*}{2} \bnabla \cdot \left( |\bnabla \psi | \bnabla \psi \right) -\nu \Delta^5 \psi \, ,  \label{eq:psireduced} \\
\partial_t  (\Delta \tau-\tau) + J(\psi, \Delta \tau - \tau) +\partial_x \psi     & = &  -\nu \Delta^{4} ( \Delta \tau - \tau)  \,  , \label{eq:taureduced}
\end{eqnarray}
where we have also included the hyperviscous terms to damp vorticity filaments and ensure numerical stability.

A few remarks are in order. Firstly, as for the initial set of equations, the reduced equations above conserve mechanical energy. The corresponding energy equation is equation~(\ref{eq:energy2LQG}) in which we substitute the (instantaneous version of) approximation~(\ref{eq:simplediss}). Secondly, as a sanity check we can estimate the rate of baroclinic-to-barotropic energy convection $\la \psi J(\tau,\Delta \tau) \ra$ induced by the term $J(\tau,\Delta \tau)$. Based on Table~\ref{table:scalings}, we obtain $\la \psi J(\tau,\Delta \tau) \ra\sim\ell_*^2$. This energy conversion rate is also proportional to the overall meridional heat flux in the system. It is therefore equal to the dimensionless diffusivity $D_*$. We thus recover the scaling relation $D_*\sim \ell_*^2$ obtained in the context of the vortex-gas scaling theory \citep{gallet2020}, while a close relation involving the energy-containing scale was discussed by~\citet{larichev1995} before that. Thirdly and more surprisingly, the equations above lack some of the linear terms that are crucial for baroclinic instability. If initialized with energetic enough initial conditions, however, a numerical simulation of equations~(\ref{eq:psireduced}-\ref{eq:taureduced}) sustains a turbulent state indefinitely, and this turbulent state closely resembles that of the full 2LQG system. To illustrate this point, we have run a suite of such numerical simulations with weak hyperviscosity, varying the drag coefficient (see section~\ref{subsection:numerics} for details on the numerical procedure). As shown in Fig. \ref{fig:Dstr_quad}, for low drag the resulting diffusivity $D_*$ agrees quantitatively with the values obtained from simulations of the full 2LQG model~(\ref{eq:psi2LQG}-\ref{eq:tau2LQG}). We thus conclude that linear instability plays no role in the energy looping pathways described by Salmon nor in setting the overall transport properties of the 2LQG system. This explains why baroclinic turbulence appears qualitatively different from simulations of the 2D Navier-Stokes equation forced by a small-scale instability mechanism \citep{van2022spontaneous}.

\section{Large-scale expansion of the Eady model\label{sec:Eadyexp}}

The next step of the mapping procedure consists of a large-scale expansion of the Eady model. To motivate this expansion, we first discuss the scale of energy conversion in our suite of numerical simulations of the 2LQG model. For brevity, we defer the discussion on numerical methods to section~\ref{subsection:numerics} where the simulations of all three models are described together.

\subsection{Preamble: the scales of energy conversion}

\begin{figure}
    \centering
\includegraphics[width=0.7\linewidth]{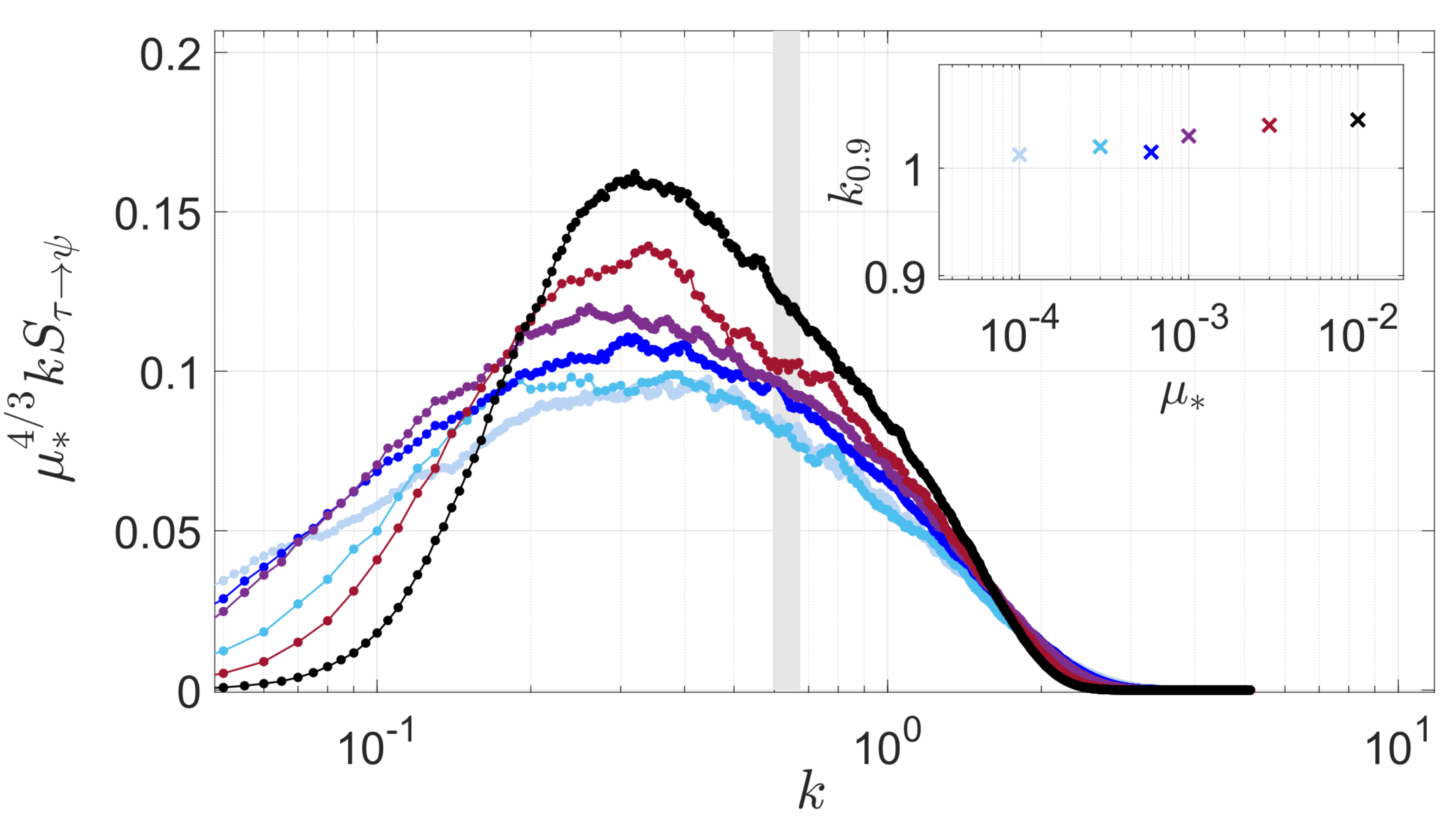}
    \caption{Premultiplied cross-spectra $S_{\tau \rightarrow \psi }(k)$ from the 2LQG model, rescaled using the power law $D_* \sim \mu_*^{-4/3}$ of the overall energy conversion rate. Color codes for the dimensionless drag $\mu_*$, with values visible in the inset. The gray-shaded interval corresponds to the range spanned by the injection wavenumber $k_\Omega$ obtained from the ratio of barotropic enstrophy injection rate to barotropic energy injection rate (see text). 
    The inset shows the wavenumber $k_{0.9}$ as a function dimensionless drag, defined such that $90$\% of energy injection into the barotropic flow arises from wavenumbers $k \leq k_{0.9}$. }
    \label{fig:cross_spt}
\end{figure}

The scale dependence of the baroclinic-to-barotropic energy conversion in the 2LQG model is investigated using the cross-spectrum $S_{\tau \to \psi}(k)$ between $\psi$ and $J(\tau, \Delta \tau)$, defined such that $\int_0^\infty S_{\tau \to \psi}(k) \mathrm{d}k= \left \langle \psi J(\tau, \Delta \tau)   \right \rangle$ (in line with the reduction in section~\ref{section:reduced_eq}, we neglect $\partial_x \Delta \tau$ as compared to the dominant term $J(\tau, \Delta \tau)$). Figure \ref{fig:cross_spt} shows the pre-multiplied cross-spectra $k \, S_{\tau \to \psi}(k)$ computed for friction coefficients ranging from $\mu_* =10^{-4}$ to $\mu_* =10^{-2}$. A key observation is that the energy transfers correspond mostly to wavenumbers $k \leq 1$, as illustrated by the fact that the pre-multiplied spectra have weight mostly for wavenumbers $k \leq 1$. 

To make this point more quantitative, we define $k_{0.9}$ as the wave number such that $\int_0^{k_{0.9}} S_{\tau \to \psi}(k) \mathrm{d}k= 0.9 \langle \psi J(\tau, \Delta \tau)\rangle$. That is, $90\%$ of energy injection into the barotropic flow is due to wavenumbers $k \leq k_{0.9}$. 
As shown in the inset of figure~\ref{fig:cross_spt} for the 2LQG simulations, $k_{0.9}$ is close to one, with an approximately constant value in the low-drag limit. Also shown in figure~\ref{fig:cross_spt} is a characteristic injection wavenumber $k_\Omega$ for the barotropic flow of the reduced model, based on the ratio of the injection rate of barotropic enstrophy to the injection rate of barotropic energy (dominant terms only):
\begin{eqnarray}
    k_\Omega=\sqrt{-\la \overline{J(\tau, \Delta \tau) \Delta \psi} \ra / \la \overline{J(\tau, \Delta \tau) \psi} \ra}.
\end{eqnarray}
As shown in figure~\ref{fig:cross_spt}, the latter is approximately constant in the low-drag regime, with $k_\Omega \simeq 0.6$.

Our strategy for establishing a mapping between the Eady and 2LQG models is the following. Based on the observation that energy injection into the barotropic mode is due to wavenumbers $k \leq 1$ in both the 2LQG model and the reduced model, we perform a low-$k$ expansion of the inversion relation~(\ref{eq:inversionEady}) of the Eady model: $k/\tanh k \simeq 1+k^2/3$ and $k/ \sinh k \simeq 1-k^2/6$ with $96\%$ accuracy for $k\leq 1.2$. In section~\ref{sec:reduction}, we show that such a low-$k$ expansion turns the Eady model into a set of equations that closely resembles the 2LQG system. The connection is made explicit by changing variables to show that the large-scale expansion of the Eady model leads to the same reduced set of equations (\ref{eq:psireduced}-\ref{eq:taureduced}) as the 2LQG model, with explicit expressions for the control parameters of the Eady model in terms of those of the 2LQG model. The reduced equations for the Eady model thus correspond to energy injection into the barotropic mode at scales $k \lesssim 1$, which justifies the low-$k$ expansion {\it a posteriori}, in a self-consistent fashion.
Reducing the two models to the same set of equations allows us to quantitatively predict the transport properties of the Eady model from the knowledge of the scaling-laws of the 2LQG model only. We validate these predictions numerically in section \ref{subsection:numerics} using simulations of both the 2LQG and the Eady model.

\subsection{Reduction of the Eady model\label{sec:reduction}}

Expanding the inversion matrix in (\ref{eq:inversionEady}) to order $k^2$ yields:
\begin{eqnarray}
\left( \begin{matrix}
\hat{b}_{0}|_{\bf k}  \\
\hat{b}_{-1}|_{\bf k} 
\end{matrix} \right) & = &
\left[ \begin{matrix}
1+ \frac{k^2}{3} &  -1+ \frac{k^2}{6}   \\
1- \frac{k^2}{6} & -1- \frac{k^2}{3} 
\end{matrix} \right]
\left( \begin{matrix}
\hat{p}_{0}|_{\bf k} \\
\hat{p}_{-1}|_{\bf k} 
\end{matrix} \right) \, , \label{eq:inversionEadyexp}
\end{eqnarray}
which we recast in physical space using the Laplacian operator $\Delta$:
\begin{eqnarray}
\left( \begin{matrix}
{b}_0 \\
{b}_{-1}
\end{matrix} \right) & = &
\left[ \begin{matrix}
1- \frac{\Delta}{3} &  -1- \frac{\Delta}{6}   \\
1+ \frac{\Delta}{6} & -1+ \frac{\Delta}{3} 
\end{matrix} \right]
\left( \begin{matrix}
{p}_0 \\
{p}_{-1}
\end{matrix} \right) \, . \label{eq:inversionEadyphysical}
\end{eqnarray}

The inversion relation above somewhat resembles that of the 2LQG system, where $b_0$ plays a role similar to $q_1$ and $- b_{-1}$ plays a role similar to $q_2$. To make this similarity visible, we change variables to: 
\begin{equation}
\tilde{{\bf x}}= \sqrt{12} \, {\bf x} \, , \qquad \tilde{\psi}=6 (p_0+p_{-1}) \, , \qquad \tilde{\tau} = \sqrt{12} (p_0 - p_{-1}) \, , \qquad \tilde{t}=t \, , \qquad \tilde{\mu}=\frac{\mu_\text{Eady}}{\sqrt{3}}\, , \label{eq:changevariables}
\end{equation}
where ${\bf x}=(x,y)$ denotes the horizontal coordinates.

Substitution into~(\ref{eq:Eadyb0}-\ref{eq:Eadybminus1}) and in the expanded inversion relation~(\ref{eq:inversionEadyphysical}) allows us to recast the governing equations in terms of the new variables. Dropping the tildes for brevity, after lengthy but straightforward calculations we obtain:

\begin{align}
%\partial_t  \Delta \psi + J(\psi,\Delta \psi) + J(\tau,\Delta \tau) + \partial_x \Delta \tau   &  = &   {\cal D} _\text{Eady} \, ,  \label{eq:mappedEadypsi} \\
%\partial_t  (\tau-\Delta \tau) + J(\psi, \tau-\Delta \tau) -  \psi_x -3 \partial_x \Delta \psi -3 J(\tau,\Delta \psi)    & = &  \sqrt{3} \, {\cal D} _\text{Eady}  \,  . \label{eq:mappedEadytau}
& \partial_t  \Delta \psi + J(\psi,\Delta \psi) + J(\tau,\Delta \tau) + \partial_x \Delta \tau     =   -\frac{\mu}{2} \bnabla \cdot \left[ |\bnabla (\psi - \sqrt{3} \tau) | \bnabla (\psi - \sqrt{3} \tau) \right] \, ,  \label{eq:mappedEadypsi} \\
& \partial_t  (\tau-\Delta \tau) + J(\psi, \tau-\Delta \tau) -  \partial_x \psi -3 \partial_x \Delta \psi -3 J(\tau,\Delta \psi)     =   - \frac{\sqrt{3} \mu}{2} \bnabla \cdot \left[ |\bnabla (\psi - \sqrt{3} \tau) | \bnabla (\psi - \sqrt{3} \tau) \right] \,  . \label{eq:mappedEadytau}
\end{align}
The last step consists in performing the same reduction of equations (\ref{eq:mappedEadypsi}-\ref{eq:mappedEadytau}) as for the 2LQG model. Namely, following the same steps as in section \ref{section:reduced_eq}, we neglect the term $\partial_x \Delta \tau$ in equation~(\ref{eq:mappedEadypsi}) and the terms $-3 \partial_x \Delta \psi -3 J(\tau,\Delta \psi)$ in equation~(\ref{eq:mappedEadytau}). Additionally, we neglect the drag term in equation~(\ref{eq:mappedEadytau}) while retaining only the (dominant) contribution from $\psi$ in the drag term of equation~(\ref{eq:mappedEadypsi}). Following these simplifications the equations~(\ref{eq:mappedEadypsi}-\ref{eq:mappedEadytau}) reduce to the exact same equations~(\ref{eq:psireduced}-\ref{eq:taureduced}) as the 2LQG model, with $\mu$ playing the role of $\mu_*$.

\begin{figure}
    \centering
    \includegraphics[width=0.8\linewidth]{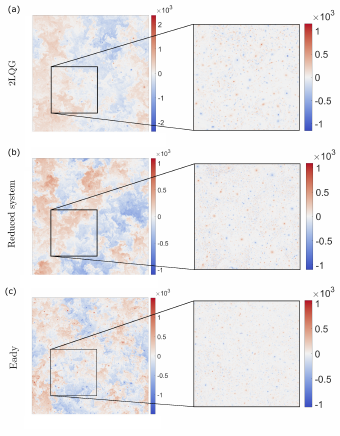}
    \caption{Snapshots from all three models. Left panels of (a) and (b) show the baroclinic streamfunction $\tau$ from the 2LQG and reduced models, to be compared with their Eady counterpart $\tilde{\tau}=\sqrt{12}(p_0-p_{-1})$, shown in the left panel of (c). The right-hand panels of (a) and (b) show the barotropic vorticity $\Delta \psi$ from the 2LQG and reduced models, to be compared with their Eady counterpart $b_0-b_{-1}$, shown in the right-hand panel of (c). The drag is $\mu_*=3 \times 10^{-4}$ in (a) and (b) and $\mu_\text{Eady}=6 \times 10^{-4}$ in (c). The Eady simulation maps onto an effective friction $\mu_\text{Eady}/\sqrt{3}\simeq 3.5 \times 10^{-4}$ for the equivalent 2LQG system, in an effective simulation domain that is larger than that of simulations (a) and (b). Hence the  smaller apparent flow scales in (c).  
    \label{fig:snapshots}}
\end{figure}

\section{Eady diffusivity inferred from the equivalent 2LQG system \label{sec:mapping}}

\subsection{Theoretical prediction}

The mapping readily provides a way of determining the effective diffusivity of the Eady model in terms of that of the 2LQG model. We thus assume that we know the functional dependence of the 2LQG effective diffusivity $D_*(\mu_*)$ with the dimensionless friction coefficient $\mu_*$. Our goal is to deduce the Eady effective diffusivity $D_\text{Eady}(\mu_\text{Eady})$ in terms of the dimensionless friction coefficient $\mu_\text{Eady}$ of the Eady model.
We start from the relation~(\ref{eq:defDeady}) under the form $D_{\text{Eady}}=\la b_0 \partial_x p_0  \ra$. Substituting expression (\ref{eq:inversionEadyphysical}) for $b_0$ in terms of $p_0$ and $p_{-1}$ leads to 
\begin{align}
D_\text{Eady} & =-\la \overline{ \partial_x (p_0) p_{-1} } \ra -\la \overline{ \partial_x (p_0) \Delta(p_{-1}) } \ra /6  \, . \label{eq:Deadytemp}
\end{align}
The change of variables~(\ref{eq:changevariables}) corresponds to $p_0=\tilde{\psi}/12+ \tilde{\tau}/(4\sqrt{3})$, $p_{-1}=\tilde{\psi}/12- \tilde{\tau}/(4\sqrt{3})$, and spatial derivatives rescaled according to $\partial_{x_i}=2\sqrt{3} \, \partial_{\tilde{x}_i}$. Substituting into~(\ref{eq:Deadytemp}) and dropping the tildes yields:
\begin{align}
D_\text{Eady} & = \frac{\la \overline{ \psi_x \tau } \ra}{12}  + \frac{\la \overline{ \psi_x \Delta \tau } \ra}{6}   \, . \label{eq:Deadytemp2}
\end{align}
According to table~\ref{table:scalings}, the term $\la \overline{ \psi_x \tau } \ra$ scales as $\ell_*^2$ at large scale and as $\ell_*^{4/3}$ at small scale, while the term $ \la \overline{ \psi_x \Delta \tau } \ra$ scales as $1$ at large scale and as $\ell_*^{4/3}$ at small scale. We conclude that $\la \overline{ \psi_x \Delta \tau } \ra \sim \ell_*^{4/3}$ is negligible as compared to $\la \overline{ \psi_x \tau } \ra \sim \ell_*^2$ on the rhs of~(\ref{eq:Deadytemp2}). To dominant order in $\ell_* \gg 1$, we finally obtain:
\begin{align}
D_\text{Eady}(\mu_{\text{Eady}}) & = \frac{1}{12} D_* \left( \mu_*={\mu_{\text{Eady}}}/{\sqrt{3}} \right) \, , \label{eq:DEady2LQG}
\end{align}
where the expression of $\mu_*$ in terms of $\mu_{\text{Eady}}$ arises from~(\ref{eq:changevariables}).
Equation~(\ref{eq:DEady2LQG}) provides a parameter-free expression of the effective diffusivity of the Eady model, based on that of the 2LQG model. A low-drag asymptotic expression for the latter is provided by the vortex-gas scaling theory \citep{hadjerci2023}:
\begin{align}
D_*(\mu_*)=\frac{c_{\text{2LQG}}}{\mu_*^{4/3}} \, ,
\end{align}
where the prefactor is $c_{\text{2LQG}}\simeq 0.3436$. Substitution into~(\ref{eq:DEady2LQG}) yields the following asymptotic prediction for the effective diffusivity of the low-drag Eady model:
\begin{align}
D_\text{Eady}(\mu_{\text{Eady}}) & = \frac{c_{\text{Eady}}}{\mu_{\text{Eady}}^{4/3}} \, , \label{eq:DEadyasymptpred}
\end{align}
with the prefactor $c_{\text{Eady}}=c_{\text{2LQG}}/(4\times 3^{1/3}) \simeq 0.060$.

\subsection{Numerical validation \label{subsection:numerics}}
\begin{figure}
    \centering
    \includegraphics[width=0.9\linewidth]{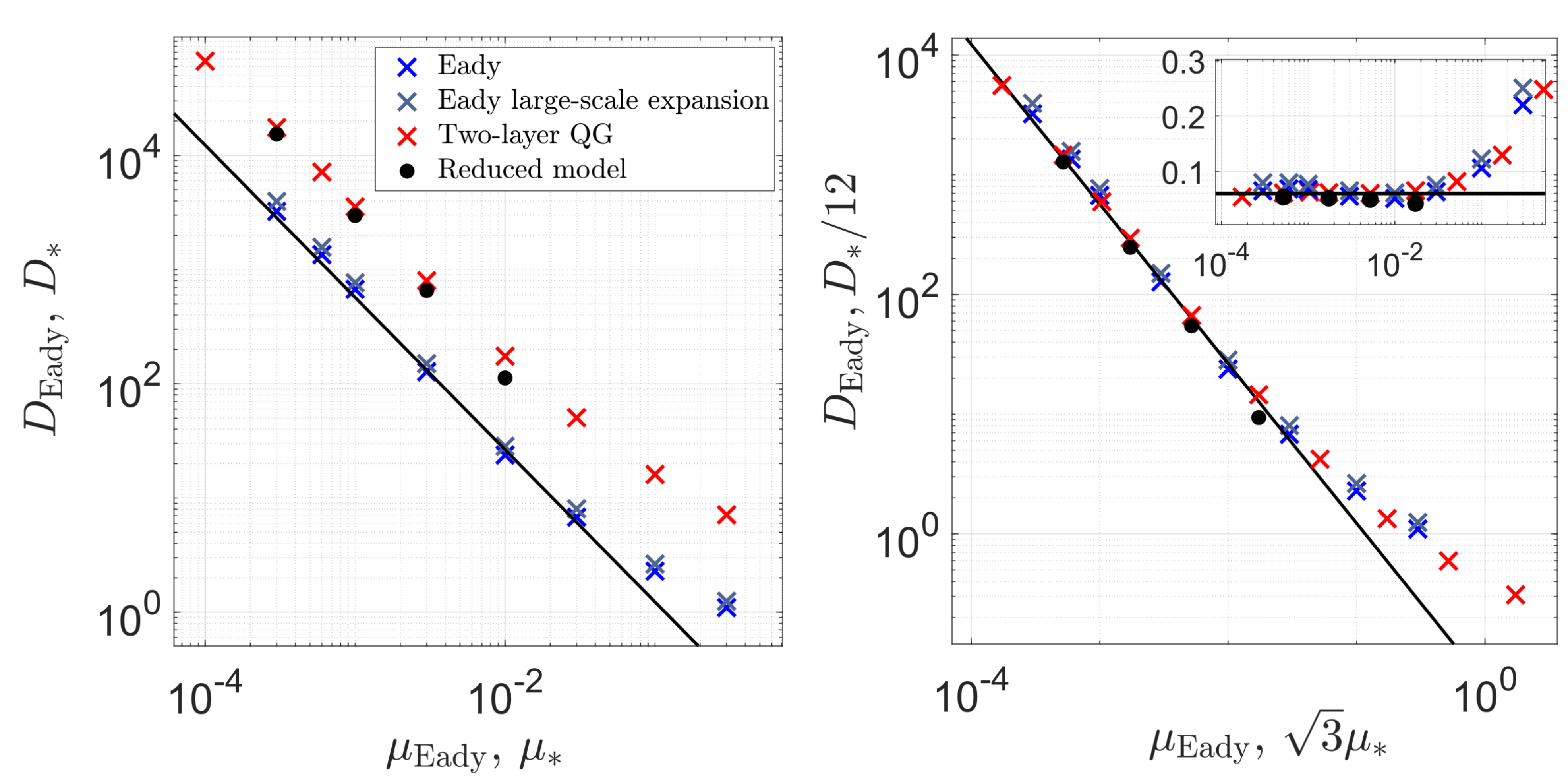}
    \caption{Effective diffusivity versus quadratic drag coefficient from numerical simulations of the Eady model (dark blue), of the expanded Eady model using  inversion relation (\ref{eq:inversionEadyexp}) (light blue), of the 2LQG model (red) and of the reduced model (\ref{eq:psireduced}-\ref{eq:taureduced}) (black circles). \textbf{Left:} raw data for all models. \textbf{Right:} The mapping suggests plotting the data from the 2LQG and reduced models under the form $D_*/12$ vs. $\sqrt{3}\mu_*$, which collapses the data of all models in the low-drag regime. The inset shows the same data compensated by a $-4/3$ power-law. 
    In each panel, the solid line represents the low-drag theoretical prediction (\ref{eq:predictionDEadylinear}) deduced from the mapping procedure.}
    \label{fig:Dstr_quad}
\end{figure}

With the goal of validating the predictions above, we performed suites of numerical simulations of the Eady model, of the 2LQG model, and of the reduced equations~(\ref{eq:psireduced}-\ref{eq:taureduced}), varying the drag coefficient for all three models. The simulations employ standard pseudo-spectral methods with dealiasing and Runge-Kutta time-stepping. 
The doubly periodic domain is large-enough for the effective diffusivity to be independent of domain size, and we employ low-enough hyperviscosity for the effective diffusivity to also be independent of  hyperviscosity. The effective diffusivity is thus a function of the drag coefficient only. Snapshots of the instantaneous fields are provided in figure~\ref{fig:snapshots} for all three models in the low-drag regime. The fields look qualitatively similar, with a barotropic flow organized into a gas of coherent vortices. At the quantitative level, the raw dataset for diffusivity versus drag is shown in figure~\ref{fig:Dstr_quad}. In the low-drag regime, the curves $D_*(\mu_*)$ are extremely similar for the 2LQG model and for the reduced equations, which confirms that the latter correctly describe the low-drag dynamics of the full 2LQG model. By contrast, the curve $D_{\text{Eady}}(\mu_{\text{Eady}})$ departs from that of the other two models. Also shown are numerical simulations of an Eady model where the inversion relation~(\ref{eq:inversionEady}) has been replaced by the large-scale expansion~(\ref{eq:inversionEadyphysical}). The resulting eddy diffusivity closely agrees with $D_\text{Eady}$, which validates the large-scale expansion of the Eady model.

To test the predictions of the mapping procedure, in Fig. \ref{fig:Dstr_quad} we plot $D_{\text{Eady}}$ vs. $\mu_{\text{Eady}}$, together with the data of the 2LQG and reduced models plotted under the form  $D_*/12$ vs. $\sqrt{3} \, \mu_*$. According to equation (\ref{eq:DEady2LQG}), we expect the data from all three models to collapse onto a master curve in the low-drag regime. Such a collapse of the whole dataset is indeed observed in Fig. \ref{fig:Dstr_quad}, which validates the mapping procedure. Also shown in Fig. \ref{fig:Dstr_quad} is the low-drag asymptotic prediction~(\ref{eq:DEadyasymptpred}) derived from the vortex-gas scaling theory. This prediction accurately captures the low-drag data, demonstrating how the 2LQG model can be used to make quantitative, parameter-free predictions for the Eady model.

\section{Linear drag\label{sec:linear}}

\begin{figure}
\centerline{\includegraphics[width=7 cm]{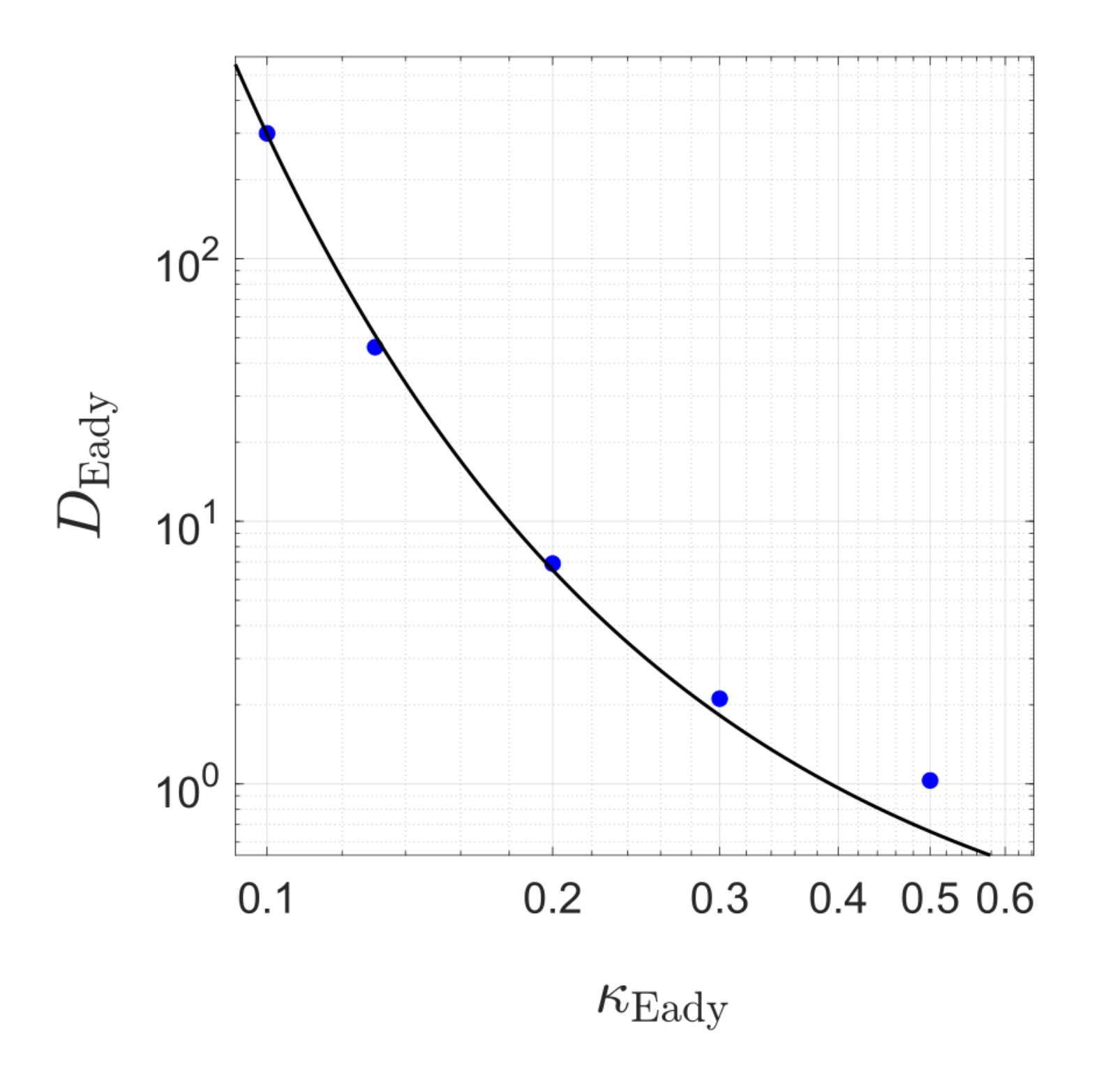}}
      \caption{Effective diffusivity versus linear drag coefficient in the Eady model. Symbols are numerical simulations. The solid line is the low-drag theoretical prediction~(\ref{eq:predictionDEadylinear}) deduced from the mapping procedure.\label{fig:lindrag}}
\end{figure}

While we have introduced the mapping procedure for an Eady model with quadratic drag, the approach readily carries over to linear bottom drag. The rhs of equation~(\ref{eq:Eadybminus1}) is then replaced by $-\kappa_\text{Eady} \Delta p_{-1}$, where $\kappa_\text{Eady}$ denotes the dimensionless linear drag coefficient (see \cite{Gallet_2022} for a detailed description of the Eady model with linear bottom drag). For a 2LQG model with linear drag, the drag term entering equations~(\ref{eq:psi2LQG}-\ref{eq:tau2LQG}) becomes ${\cal D}=-\kappa_* \Delta(\psi-\tau)$, where $\kappa_*$ denotes the linear drag coefficient non-dimensionalized using $U$ and $\lambda$. Following the steps in section \ref{sec:mapping}, one obtains the reduced set of equations~(\ref{eq:psireduced}-\ref{eq:taureduced}), where the drag term is replaced by $-\kappa_* \Delta \psi$. Introducing the rescaled variables~(\ref{eq:changevariables}) with the additional relation $\tilde{\kappa}=\kappa_\text{Eady}$, one easily shows that the Eady model reduces to the same reduced equations as the 2LQG model. This mapping provides the following relation between the effective diffusivities of the Eady and 2LQG models with weak linear drag:
\begin{align}
D_\text{Eady}(\kappa_{\text{Eady}}) & = \frac{1}{12} D_* \left( \kappa_*=\kappa_{\text{Eady}} \right) \, . \label{eq:DEady2LQGlinear}
\end{align}
The vortex-gas theory applied to the 2LQG model leads to the scaling prediction $D_*(\kappa_*)=c_1\,\exp(c_2/\kappa_*)$ for the effective diffusivity, where the numerical constants $c_1=1.7128$ and $c_2=0.7644$ have been adjusted to the numerical data in~\citet{hadjerci2023}. Combining the knowledge of the 2LQG diffusivity with the mapping relation (\ref{eq:DEady2LQGlinear}) leads to a quantitative prediction for the effective diffusivity of the Eady model with weak linear drag:
\begin{align}
D_\text{Eady}(\kappa_{\text{Eady}}) & =  \frac{c_1}{12} e^{c_2/\kappa_{\text{Eady}}} \simeq  0.1427 e^{0.7644/\kappa_{\text{Eady}}} \, .\label{eq:predictionDEadylinear}
\end{align}

In figure~\ref{fig:lindrag}, we compare this prediction with numerical simulations of the QG Eady model with linear bottom drag. The numerical data correspond to the QG Eady simulations reported in~\citet{Gallet_2022}, where we ran again the lowest-drag simulation using a larger domain and a smaller hyperviscosity to better converge the diffusivity. The low-drag numerical data in figure~\ref{fig:lindrag} agree quantitatively with the theoretical prediction~(\ref{eq:predictionDEadylinear}), which confirms that the mapping procedure is valid for both quadratic and linear drag.

\section{Conclusion\label{sec:conclusion}}

Despite the different nature of the PV gradients involved in the Eady model and in the 2LQG model, the present study shows that the two systems share the exact same quantitative behavior at large scale. Indeed, the two models asymptote to the same `reduced' set of equations in the limit of weak bottom drag, that is, when the integral scale of the flow is large compared to the Rossby deformation radius. The heat transport of the Eady model is thus depth-invariant, with a magnitude governed by an equivalent 2LQG model. As an illustration of this procedure, we derived a quantitative prediction for the low-drag asymptotic behavior of the Eady diffusivity based on results from a vortex-gas scaling theory developed for the 2LQG model. Of course, the mapping between the two models holds for large-scale quantities only, such as the eddy diffusivity. By contrast, flow structures at scales smaller than the deformation radius strongly differ between the 2LQG and the Eady model, as only the latter model presents the characteristics of surface quasi-geostrophy (e.g. fronts) near top and bottom \citep{Lapeyre2017}.

A surprising aspect of the reduced set of equations is that, while capturing large-scale 2LQG and Eady turbulence quantitatively, it does not capture linear baroclinic instability. This indicates that the properties of the turbulent flow are, in fact, somewhat disconnected from the instability mechanism, questioning parameterizations where the turbulent (Gent-McWilliams) diffusivity is assumed to be controlled by the Eady growth rate \citep{visbeck1997, Treguier1997}.

A natural extension of the present study would be to consider systems with a nonzero background PV gradient~\citep{miller2024gyre,miller2025impact}, resulting from latitudinal variations of the Coriolis parameter, sloping bottom topography~\citep{deng2024distinct} or vertically dependent background shear and stratification~\citep{smith2009evidence,meunier2023}. The first step in this direction could be a Charney model with weak planetary $\beta$~\citep{charney1947dynamics,green1960problem}. In the weak-$\beta$ regime, we derived a perturbative prediction for the vertical profile of the turbulent heat flux throughout the water column, up to an unknown prefactor set by the overall (depth-integrated) heat flux. An extension of the present mapping to $\beta \neq 0$ would complement the vertical profile with a prediction for the depth-integrated heat flux, based on the behavior of a two-layer model with weak $\beta$~\citep{gallet2021quantitative}.

\backsection[Funding]{This research is supported by the European Research Council under grant agreement 101124590.}

\backsection[Declaration of interests]{ The authors report no conflict of interest.}

\bibliographystyle{jfm}
\bibliography{main}

@article{arbicfierl2004a,
  title={Effects of mean flow direction on energy, isotropy, and coherence of baroclinically unstable beta-plane geostrophic turbulence},
  author={Arbic, Brian K and Flierl, Glenn R},
  journal={Journal of physical oceanography},
  volume={34},
  number={1},
  pages={77--93},
  year={2004}
}

@article{chang_parameter_2021,
	title = {The {Parameter} {Dependence} of {Eddy} {Heat} {Flux} in a {Homogeneous} {Quasigeostrophic} {Two}-{Layer} {Model} on a beta {Plane} with {Quadratic} {Friction}},
	volume = {78},
	issn = {0022-4928, 1520-0469},
	url = {https://journals.ametsoc.org/view/journals/atsc/78/1/jas-d-20-0145.1.xml},
	doi = {10.1175/JAS-D-20-0145.1},
	number = {1},
	urldate = {2023-08-01},
	journal = {Journal of the Atmospheric Sciences},
	author = {Chang, Chiung-Yin and Held, Isaac M.},
	month = jan,
	year = {2021},
	pages = {97--106},
}

@article{chang_control_2019,
	title = {The {Control} of {Surface} {Friction} on the {Scales} of {Baroclinic} {Eddies} in a {Homogeneous} {Quasigeostrophic} {Two}-{Layer} {Model}},
	volume = {76},
	issn = {0022-4928, 1520-0469},
	url = {https://journals.ametsoc.org/view/journals/atsc/76/6/jas-d-18-0333.1.xml},
	doi = {10.1175/JAS-D-18-0333.1},
	number = {6},
	urldate = {2023-08-01},
	journal = {Journal of the Atmospheric Sciences},
	author = {Chang, Chiung-Yin and Held, Isaac M.},
	month = jun,
	year = {2019},
	pages = {1627--1643},
}

@article{charney1947dynamics,
  title={The dynamics of long waves in a baroclinic westerly current},
  author={Charney, Jule G},
  journal={Journal of Atmospheric Sciences},
  volume={4},
  number={5},
  pages={136--162},
  year={1947}
}

@article{chen_revisiting_2023,
	title = {Revisiting the {Baroclinic} {Eddy} {Scalings} in {Two}-{Layer}, {Quasigeostrophic} {Turbulence}: {Effects} of {Partial} {Barotropization}},
	volume = {53},
	issn = {0022-3670, 1520-0485},
	shorttitle = {Revisiting the {Baroclinic} {Eddy} {Scalings} in {Two}-{Layer}, {Quasigeostrophic} {Turbulence}},
	url = {https://journals.ametsoc.org/view/journals/phoc/53/3/JPO-D-22-0102.1.xml},
	doi = {10.1175/JPO-D-22-0102.1},
	number = {3},
	urldate = {2023-08-01},
	journal = {Journal of Physical Oceanography},
	author = {Chen, Shih-Nan},
	month = mar,
	year = {2023},
	pages = {891--913},
}

@article{constantin1994formation,
  title={Formation of strong fronts in the 2-D quasigeostrophic thermal active scalar},
  author={Constantin, Peter and Majda, Andrew J and Tabak, Esteban},
  journal={Nonlinearity},
  volume={7},
  number={6},
  pages={1495},
  year={1994},
  publisher={IOP Publishing}
}

@article{deng2024distinct,
  title={Distinct impacts of topographic versus planetary PV gradients on baroclinic turbulence},
  author={Deng, Peng and Wang, Yan},
  journal={Journal of Physical Oceanography},
  volume={54},
  number={10},
  pages={2205--2231},
  year={2024},
  publisher={American Meteorological Society}
}

@article{eady1949,
  title={Long waves and cyclone waves},
  author={Eady, Eric T},
  journal={Tellus},
  volume={1},
  number={3},
  pages={33--52},
  year={1949},
  publisher={Taylor \& Francis}
}

@article{flierl1978,
  title={Models of vertical structure and the calibration of two-layer models},
  author={Flierl, Glenn R},
  journal={Dynamics of Atmospheres and Oceans},
  volume={2},
  number={4},
  pages={341--381},
  year={1978},
  publisher={Elsevier}
}

@article{gallet2020,
  title={The vortex gas scaling regime of baroclinic turbulence},
  author={Gallet, Basile and Ferrari, Raffaele},
  journal={Proceedings of the National Academy of Sciences},
  volume={117},
  number={9},
  pages={4491--4497},
  year={2020},
  publisher={National Academy of Sciences}
}

@article{gallet2021quantitative,
  title={A quantitative scaling theory for meridional heat transport in planetary atmospheres and oceans},
  author={Gallet, Basile and Ferrari, Raffaele},
  journal={AGU Advances},
  volume={2},
  number={3},
  pages={e2020AV000362},
  year={2021},
  publisher={Wiley Online Library}
}

@Article{Gallet_2022,
   title={Transport and emergent stratification in the equilibrated Eady model: the vortex-gas scaling regime},
   volume={948},
   ISSN={1469-7645},
   url={http://dx.doi.org/10.1017/jfm.2022.501},
   DOI={10.1017/jfm.2022.501},
   journal={Journal of Fluid Mechanics},
   publisher={Cambridge University Press (CUP)},
   author={Gallet, Basile and Miquel, Benjamin and Hadjerci, Gabriel and Burns, Keaton J. and Flierl, Glenn R. and Ferrari, Raffaele},
   year={2022},
   month=sep }

@article{green1960problem,
  title={A problem in baroclinic stability},
  author={Green, JSA},
  journal={Quarterly Journal of the Royal Meteorological Society},
  volume={86},
  number={368},
  pages={237--251},
  year={1960},
  publisher={Wiley Online Library}
}

@article{hadjerci2023,
  title={Vortex core radius in baroclinic turbulence: Implications for scaling predictions},
  author={Hadjerci, Gabriel and Gallet, Basile},
  journal={Physical Review Fluids},
  volume={8},
  number={9},
  pages={094501},
  year={2023},
  publisher={APS}
}

@article{kohl2022diabatic,
  title={The diabatic Rossby vortex: Growth rate, length scale, and the wave--vortex transition},
  author={Kohl, Matthieu and O’Gorman, Paul A},
  journal={Journal of the Atmospheric Sciences},
  volume={79},
  number={10},
  pages={2739--2755},
  year={2022}
}

@article{lacasce2012surface,
  title={Surface quasigeostrophic solutions and baroclinic modes with exponential stratification},
  author={LaCasce, JH},
  journal={Journal of Physical Oceanography},
  volume={42},
  number={4},
  pages={569--580},
  year={2012}
}

@article{lapeyre2009vertical,
  title={What vertical mode does the altimeter reflect? On the decomposition in baroclinic modes and on a surface-trapped mode},
  author={Lapeyre, Guillaume},
  journal={Journal of Physical Oceanography},
  volume={39},
  number={11},
  pages={2857--2874},
  year={2009}
}

@Article{Lapeyre2017,
AUTHOR = {Lapeyre, Guillaume},
TITLE = {Surface Quasi-Geostrophy},
JOURNAL = {Fluids},
VOLUME = {2},
YEAR = {2017},
NUMBER = {1},
ARTICLE-NUMBER = {7},
URL = {https://www.mdpi.com/2311-5521/2/1/7},
ISSN = {2311-5521},
DOI = {10.3390/fluids2010007}
}

@article{larichev1995,
  title={Eddy amplitudes and fluxes in a homogeneous model of fully developed baroclinic instability},
  author={Larichev, Vitaly D and Held, Isaac M},
  journal={Journal of physical oceanography},
  volume={25},
  number={10},
  pages={2285--2297},
  year={1995}
}

@article{meunier2023,
  title={A direct derivation of the Gent--McWilliams/Redi diffusion tensor from quasi-geostrophic dynamics},
  author={Meunier, Julie and Miquel, Benjamin and Gallet, Basile},
  journal={Journal of Fluid Mechanics},
  volume={963},
  pages={A22},
  year={2023},
  publisher={Cambridge University Press}
}

@article{GRLMeunier,
author = {Meunier, Julie and Miquel, Benjamin and Gallet, Basile},
title = {Vertical Structure of Buoyancy Transport by Ocean Baroclinic Turbulence},
journal = {Geophysical Research Letters},
volume = {50},
number = {17},
pages = {e2023GL103948},
doi = {https://doi.org/10.1029/2023GL103948},
url = {https://agupubs.onlinelibrary.wiley.com/doi/abs/10.1029/2023GL103948},
eprint = {https://agupubs.onlinelibrary.wiley.com/doi/pdf/10.1029/2023GL103948},
note = {e2023GL103948 2023GL103948},
year = {2023}
}

@article{meunier2025,
  title = {Effective Transport by 2D Turbulence: Vortex-Gas Theory vs Scale-Invariant Inverse Cascade},
  author = {Meunier, Julie and Gallet, Basile},
  journal = {Phys. Rev. Lett.},
  volume = {134},
  issue = {7},
  pages = {074101},
  numpages = {7},
  year = {2025},
  month = {Feb},
  publisher = {American Physical Society},
  doi = {10.1103/PhysRevLett.134.074101},
  url = {https://link.aps.org/doi/10.1103/PhysRevLett.134.074101}
}

@article{miller2024gyre,
  title={Gyre turbulence: Anomalous dissipation in a two-dimensional ocean model},
  author={Miller, Lennard and Deremble, Bruno and Venaille, Antoine},
  journal={Physical Review Fluids},
  volume={9},
  number={5},
  pages={L051801},
  year={2024},
  publisher={APS}
}

@article{miller2025impact,
  title={The Impact of Stratification on Surface-Intensified Eastward Jets in Turbulent Gyres},
  author={Miller, Lennard and Deremble, Bruno and Venaille, Antoine},
  journal={Journal of Physical Oceanography},
  year={2025},
  publisher={American Meteorological Society}
}

@book{pedlosky2013,
  title={Geophysical fluid dynamics},
  author={Pedlosky, Joseph},
  year={2013},
  publisher={Springer Science \& Business Media}
}

@article{phillips1954,
  title={Energy transformations and meridional circulations associated with simple baroclinic waves in a two-level, quasi-geostrophic model},
  author={Phillips, Norman A},
  journal={Tellus},
  volume={6},
  number={3},
  pages={274--286},
  year={1954},
  publisher={Taylor \& Francis}
}

@article{pudig2025baroclinic,
  title={Baroclinic Turbulence Above Rough Topography: The Vortex Gas and Topographic Turbulence Regimes},
  author={Pudig, MP and Smith, K Shafer},
  journal={Journal of Physical Oceanography},
  volume={55},
  number={5},
  pages={611--630},
  year={2025},
  publisher={American Meteorological Society}
}

@article{rocha2016galerkin,
  title={On Galerkin approximations of the surface active quasigeostrophic equations},
  author={Rocha, Cesar B and Young, William R and Grooms, Ian},
  journal={Journal of Physical Oceanography},
  volume={46},
  number={1},
  pages={125--139},
  year={2016}
}

@article{salmon1978two,
  title={Two-layer quasi-geostrophic turbulence in a simple special case},
  author={Salmon, Rick},
  journal={Geophysical \& Astrophysical Fluid Dynamics},
  volume={10},
  number={1},
  pages={25--52},
  year={1978},
  publisher={Taylor \& Francis}
}

@article{salmon1980baroclinic,
  title={Baroclinic instability and geostrophic turbulence},
  author={Salmon, Rick},
  journal={Geophysical \& Astrophysical Fluid Dynamics},
  volume={15},
  number={1},
  pages={167--211},
  year={1980},
  publisher={Taylor \& Francis}
}

@book{salmonbook,
  title={Lectures on geophysical fluid dynamics},
  author={Salmon, Rick},
  year={1998},
  publisher={Oxford University Press, USA}
}

@article{scott2012assessment,
  title={Assessment of traditional and new eigenfunction bases applied to extrapolation of surface geostrophic current time series to below the surface in an idealized primitive equation simulation},
  author={Scott, Robert B and Furnival, Darran G},
  journal={Journal of physical oceanography},
  volume={42},
  number={1},
  pages={165--178},
  year={2012}
}

@article{scott2014numerical,
  title={Numerical simulation of a self-similar cascade of filament instabilities in the surface quasigeostrophic system},
  author={Scott, Richard Kirkness and Dritschel, DG},
  journal={Physical Review Letters},
  volume={112},
  number={14},
  pages={144505},
  year={2014},
  publisher={APS}
}

@article{smith2013surface,
  title={A surface-aware projection basis for quasigeostrophic flow},
  author={Smith, K Shafer and Vanneste, Jacques},
  journal={Journal of Physical Oceanography},
  volume={43},
  number={3},
  pages={548--562},
  year={2013}
}

@article{smith2009evidence,
  title={Evidence for enhanced eddy mixing at middepth in the Southern Ocean},
  author={Smith, K Shafer and Marshall, John},
  journal={Journal of Physical Oceanography},
  volume={39},
  number={1},
  pages={50--69},
  year={2009}
}

@article{sterl2025joint,
  title={The joint effects of planetary, topography and friction on baroclinic instability in a two-layer quasi-geostrophic model},
  author={Sterl, Miriam F and Pal{\'o}czy, Andr{\'e} and Groeskamp, Sjoerd and Baatsen, Michiel LJ and LaCasce, Joseph H and Isachsen, P{\aa}l Erik},
  journal={Journal of Fluid Mechanics},
  volume={1012},
  pages={A1},
  year={2025},
  publisher={Cambridge University Press}
}

@article{thompsonyoung2006,
      author = "Andrew F. Thompson and William R. Young",
      title = "Scaling Baroclinic Eddy Fluxes: Vortices and Energy Balance",
      journal = "Journal of Physical Oceanography",
      year = "2006",
      publisher = "American Meteorological Society",
      address = "Boston MA, USA",
      volume = "36",
      number = "4",
      doi = "10.1175/JPO2874.1",
      pages=      "720 - 738",
      url = "https://journals.ametsoc.org/view/journals/phoc/36/4/jpo2874.1.xml"
}

@article {Treguier1997,
      author = "A. M. Treguier and I. M. Held and V. D. Larichev",
      title = "Parameterization of Quasigeostrophic Eddies in Primitive Equation Ocean Models",
      journal = "Journal of Physical Oceanography",
      year = "1997",
      publisher = "American Meteorological Society",
      address = "Boston MA, USA",
      volume = "27",
      number = "4",
      doi = "10.1175/1520-0485(1997)027<0567:POQEIP>2.0.CO;2",
      pages=      "567 - 580",
      url = "https://journals.ametsoc.org/view/journals/phoc/27/4/1520-0485_1997_027_0567_poqeip_2.0.co_2.xml"
}

@article{tulloch2009quasigeostrophic,
  title={Quasigeostrophic turbulence with explicit surface dynamics: Application to the atmospheric energy spectrum},
  author={Tulloch, Ross and Smith, K Shafer},
  journal={Journal of the Atmospheric Sciences},
  volume={66},
  number={2},
  pages={450--467},
  year={2009}
}

@article{valade2025surface,
  title={Surface quasigeostrophic turbulence: The refined study of an active scalar},
  author={Valade, Nicolas and Bec, J{\'e}r{\'e}mie and Thalabard, Simon},
  journal={arXiv preprint arXiv:2503.16294},
  year={2025}
}

@book{vallis2017atmospheric,
  title={Atmospheric and oceanic fluid dynamics},
  author={G. K. Vallis},
  year={2017},
  publisher={Cambridge University Press}
}

@article{van2022spontaneous,
  title={Spontaneous suppression of inverse energy cascade in instability-driven 2-D turbulence},
  author={van Kan, Adrian and Favier, Benjamin and Julien, Keith and Knobloch, Edgar},
  journal={Journal of Fluid Mechanics},
  volume={952},
  pages={R4},
  year={2022},
  publisher={Cambridge University Press}
}

@article{visbeck1997,
  title={Specification of eddy transfer coefficients in coarse-resolution ocean circulation models},
  author={Visbeck, Martin and Marshall, John and Haine, Tom and Spall, Mike},
  journal={Journal of physical oceanography},
  volume={27},
  number={3},
  pages={381--402},
  year={1997}
}

@article{yassin2022discrete,
  title={On the discrete normal modes of quasigeostrophic theory},
  author={Yassin, Houssam and Griffies, Stephen M},
  journal={Journal of Physical Oceanography},
  volume={52},
  number={2},
  pages={243--259},
  year={2022}
}

\end{document}